\documentclass[iop]{emulateapj}

\renewcommand{\deg}{^{\circ}}

\newcommand{\xic}{\xi_{cr}}

\newcommand{\m}{\mathcal M}

\usepackage[colorlinks,urlcolor=blue,citecolor=blue,linkcolor=blue]{hyperref}

\shorttitle{Shock acceleration: magnetic field amplification}
\shortauthors{Caprioli \& Spitkovsky}

\begin{document}

\title{Simulations of Ion Acceleration at Non-relativistic Shocks.\\
II. Magnetic Field Amplification}

\author{D. Caprioli and A. Spitkovsky}
\affil{Department of Astrophysical Sciences, Princeton University, 
    4 Ivy Ln., Princeton NJ 08544}
\email{caprioli@astro.princeton.edu}

\begin{abstract}
We use large hybrid simulations to study ion acceleration and generation of magnetic turbulence due to the streaming of particles that are self-consistently accelerated at non-relativistic shocks.  When acceleration is efficient, we find that the upstream magnetic field is significantly amplified.  The total amplification factor is larger than 10 for shocks with Alfv\'enic Mach number $M=100$, and scales with the square root of $M$. 
The spectral energy density of excited magnetic turbulence is determined by the energy distribution of accelerated particles, and for moderately-strong shocks ($M\lesssim30$) agrees well with the prediction of resonant streaming instability, in the framework of quasilinear theory of diffusive shock acceleration.
For $M\gtrsim30$, instead, Bell's non-resonant hybrid (NRH) instability is predicted and found to grow faster than resonant instability.
NRH modes are excited far upstream by escaping particles, and initially grow without disrupting the current, their typical wavelengths being much shorter than the current ions' gyroradii.
Then, in the nonlinear stage, most unstable modes migrate to larger and larger wavelengths, eventually becoming resonant in wavelength with the driving ions, which start diffuse.
Ahead of strong shocks we distinguish two regions, separated by the free-escape boundary: the far upstream, where field amplification is provided by the current of escaping ions via NRH instability, and the shock precursor, where energetic particles are effectively magnetized, and field amplification is provided by the current in diffusing ions. 
The presented scalings of magnetic field amplification enable the inclusion of self-consistent microphysics into phenomenological models of ion acceleration at non-relativistic shocks.
\end{abstract}


\section{Introduction}

Strong astrophysical shocks are often associated with non-thermal emission and with magnetic field amplification.
This evidence  suggests that shocks are sites of efficient particle acceleration, and that this overdensity of energetic particles is responsible for the excitation of magnetic turbulence via plasma instabilities.

This paper is the second in a series of works that study different aspects of particle acceleration at non-relativistic collisionless shocks by means of self-consistent kinetic simulations.
In particular, we use unprecedentedly large hybrid (kinetic ions--fluid electrons) simulations of high-Mach-number shocks to investigate the strongly non-linear interplay between accelerated particles and the electromagnetic field. 

In the previous paper, \citeauthor{DSA} \citeyearpar[][hereafter Paper I]{DSA}, we showed that ion acceleration can be very efficient (10--20\% of the bulk flow energy channeled in energetic particles), especially at parallel and quasi-parallel shocks, i.e., shocks propagating almost in the direction of the background magnetic field, ${\bf B}_0$. 
Quasi-parallel shocks also show an effective amplification of the initial magnetic field due to the current of energetic ions that propagate anisotropically into the upstream.
Also, 2D and 3D hybrid simulations with large computational boxes in the transverse direction revealed the formation of upstream filaments and cavities, which eventually trigger the Richtmyer--Meshkov instability at the shock, and lead to further turbulent magnetic field amplification in the downstream region \citep[see][hereafter CS13]{filam}.

The most prominent observational evidence of magnetic field amplification at strong shocks is found at the blast waves of Supernova Remnants (SNRs).
Observed strength, variability, and morphology of the synchrotron emission produced by relativistic electrons suggest that in young SNRs magnetic fields are 50--100 times larger than in the interstellar medium \citep[see, e.g.,][for diverse observational facts]{P+06,Uchiyama+07,tycho,reynoso+13}.
High-resolution X-ray images of SN1006 also indicate that magnetic field amplification must occur in the shock precursor, and not downstream \citep[see][]{gio-sn1006}. 

The main goal of this work is to use kinetic simulations to study how particles energized via \emph{diffusive shock acceleration} \citep[DSA, e.g.,][]{bell78a,blandford-ostriker78} induce magnetic field amplification in non-relativistic collisionless shocks.
The back-reaction of such self-generated magnetic turbulence on particle scattering is presented in \citep[][hereafter, Paper III]{diffusion}.
The present paper is structured as follows. 
In section \ref{sec:hybrid} we present the hybrid technique, along with some typical simulation outputs. 
Section \ref{sec:MFA} provides a description of magnetic field amplification for shocks with different strengths.
The spectrum of the self-generated magnetic turbulence is discussed in section \ref{sec:turb}, and in section \ref{sec:NRH} we outline the role of the non-resonant instability \citep{bell04}.  
We conclude in section \ref{sec:concl}.

\section{Hybrid simulations}\label{sec:hybrid}
\begin{table}
  \caption{Parameters of the hybrid runs in the paper}
  \label{tab:box}
  \begin{center}
    \begin{tabular}{cccccc} \hline \hline              
  Run & M & $x$ $[c/\omega_p]$  & $y$ $[c/\omega_p]$ & $t_{max}[\omega_c^{-1}]$ & $\Delta t [\omega_c^{-1}]$\\ 
  \hline 
  A	  & 20 & $5\times 10^4$	& $1000$ & $1000$ & $5\times 10^{-4}$ \\
  B	  & 20 & $ 10^5$	& $100$ & $2500$ & $5\times 10^{-4}$  \\
  C	  & 100 & $ 3\times 10^4$ & $2000$ & $200$ & $ 10^{-4}$  \\
  D 	  & 80 & $ 4\times 10^5$ & $200$ & $500$ & $ 2.5\times 10^{-4}$  \\
  E	  & $10\to 50$ & $ 2\times 10^4$ & $500$ & $200$ & $ 10^{-2}/M$  \\
  \hline
    \end{tabular}
  \end{center}
\end{table}

\begin{figure*}\centering
\includegraphics[trim=0px 0px 0px 0px, clip=true, width=\textwidth]{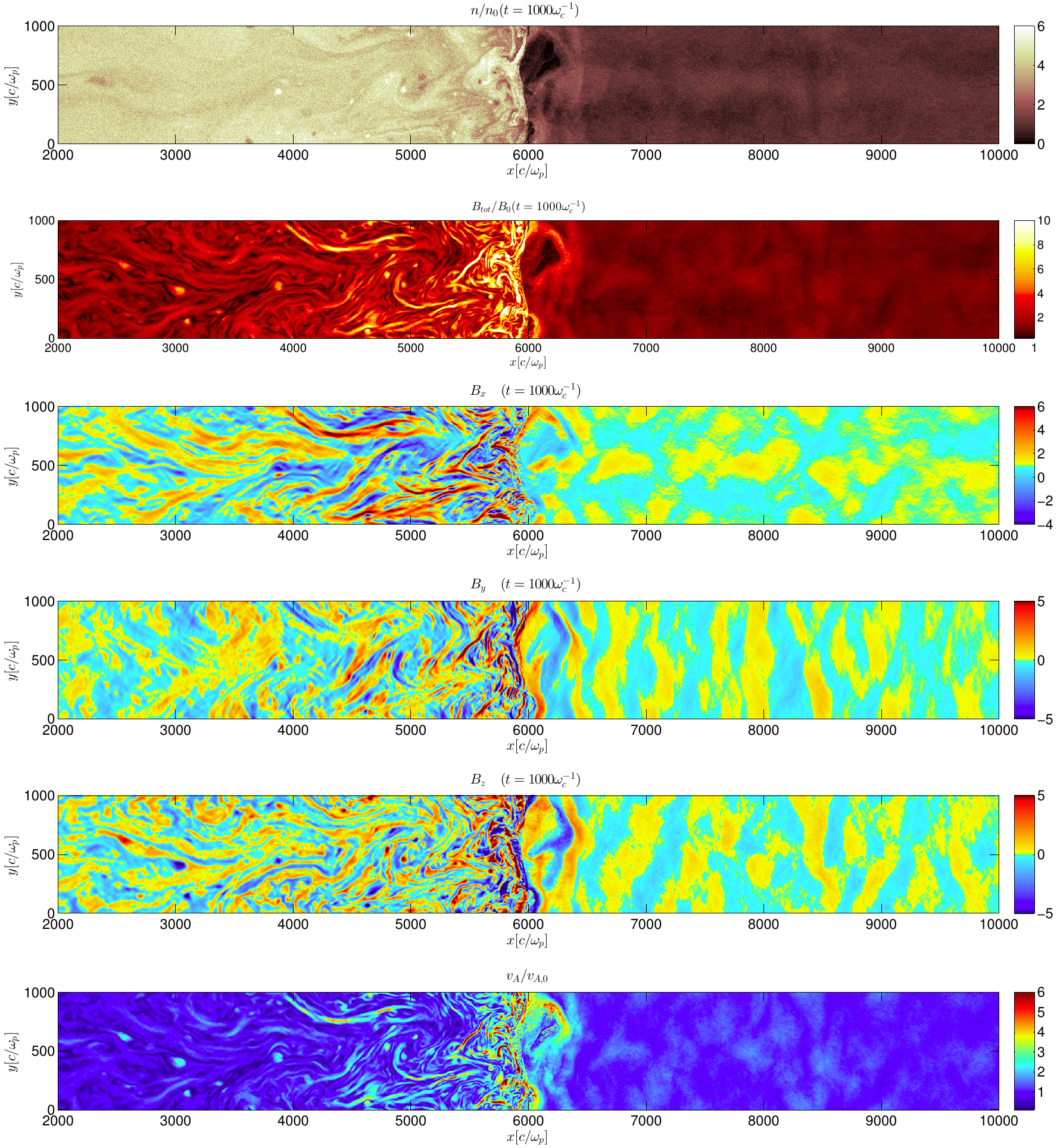}
\caption{\label{fig:summary}
Relevant physical quantities for a parallel shock with $M=20$ at $t=1000\omega_c^{-1}$ (Run A in Table \ref{tab:box}). 
From top to bottom: ion density, modulus and three components of the magnetic field, and local Alfv\'en velocity $v_A=|{\bf B}|/\sqrt{4\pi mn}$, in units of their respective initial values.
Only a portion of the computational box is shown, to emphasize the shock transition.
\emph{A color figure is available in the online journal.}}
\end{figure*}

\begin{figure}\centering
\includegraphics[trim=10px 20px 0px 320px, clip=true, width=.5\textwidth]{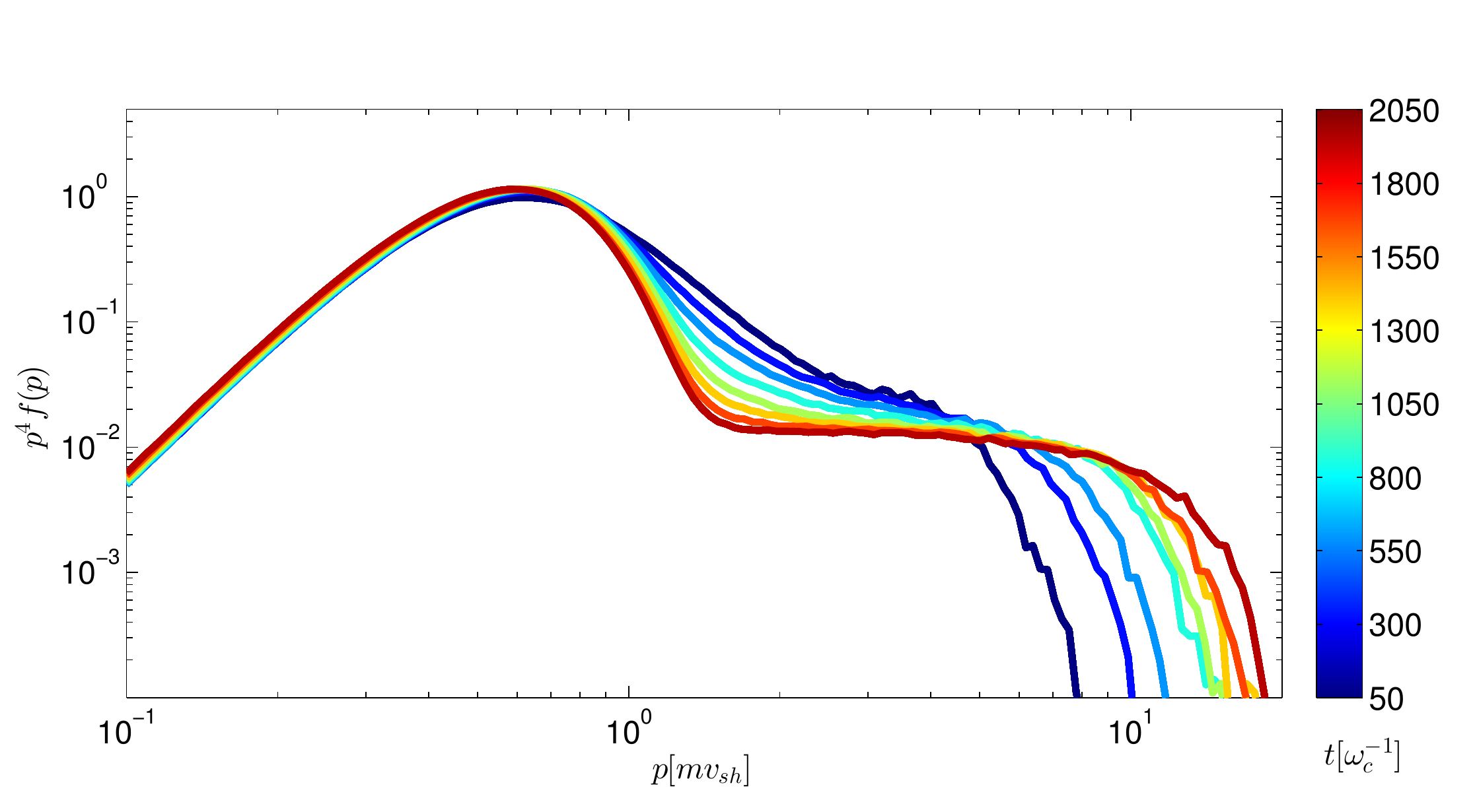}
\caption{\label{fig:evo}
Time evolution of the downstream ion momentum spectrum for $M=20$ parallel shock (Run B, see Table \ref{tab:box}), showing both the thermal component ($p\lesssim 2mv_{sh}$), and accelerated particles. 
The non-thermal power-law tail $\propto p^{-4}$ agrees with DSA prediction at strong shocks (see Paper I).
The maximum momentum increases until $t\approx 2000\omega_c^{-1}$, when the diffusion length of the most energetic ions becomes comparable with the box size.
}
\end{figure}

\begin{figure*}\centering
\includegraphics[trim=0px 0px 0px 0px, clip=true, width=\textwidth]{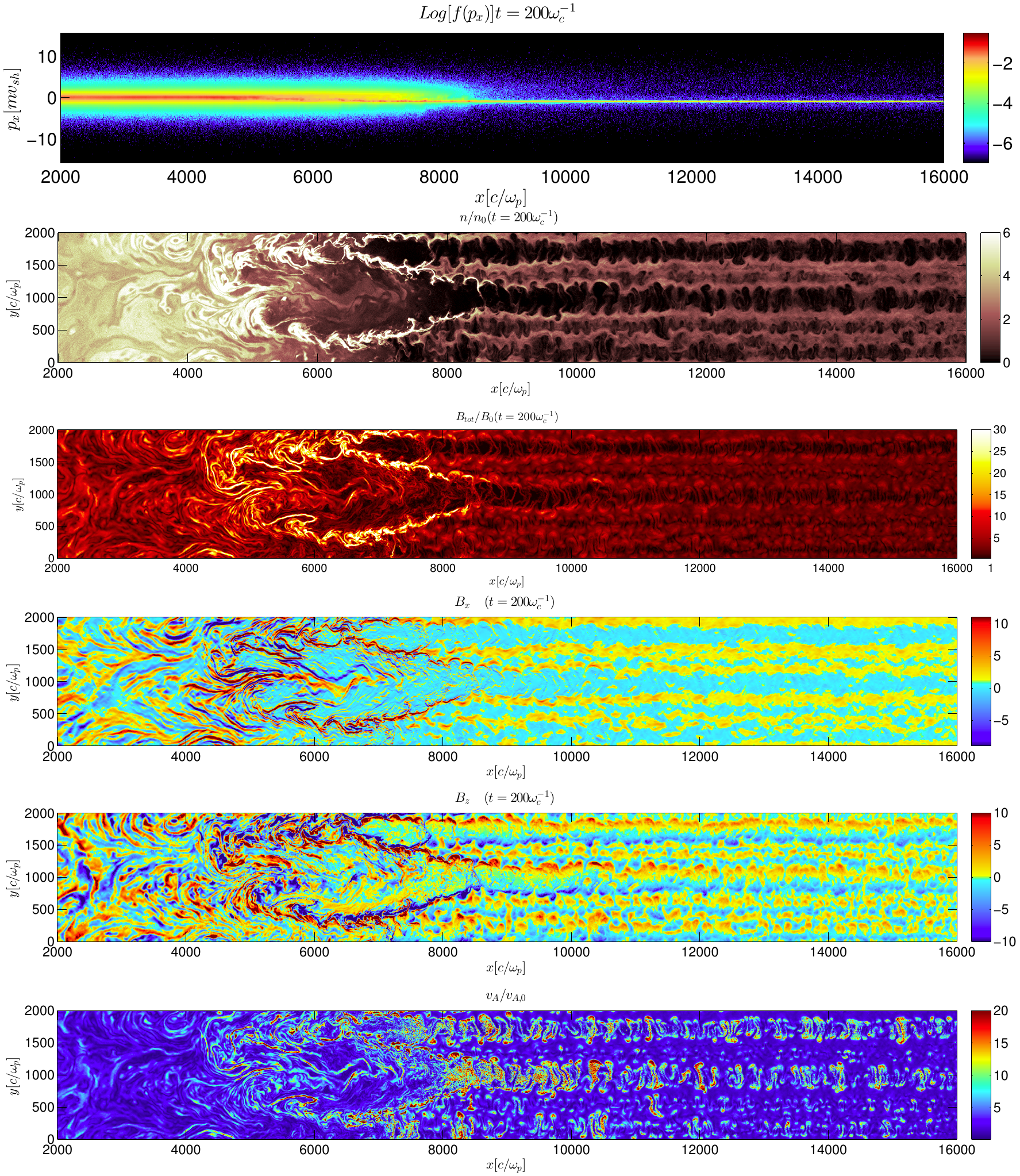}
\caption{\label{fig:summary100}
Relevant physical quantities for a parallel shock with $M=100$ at $t=200\omega_c^{-1}$, as a function of $x$ (Run C in Table \ref{tab:box}). 
From top to bottom: parallel component of the ion momentum, ion density, total magnetic field, parallel and out of plane components of the magnetic field, and Alfv\'en velocity.
\emph{A color figure is available in the online journal.}}
\end{figure*}

The simulations presented here have been performed with \emph{dHybrid}, a massively-parallel, non-relativistic, hybrid code \citep{gargate+07}.
In the hybrid limit, ions are treated kinetically, while electrons are assumed to be a neutralizing fluid with a polytropic equation of state.
Even in 2D setups, the three components of the ion momentum and of the electromagnetic field are retained.

Lengths are measured in units of the ion skin depth $c/\omega_p$, where $c$ is the light speed and $\omega_p=\sqrt{4\pi n e^2/m}$ is the ion plasma frequency, with $m,e$ and $n$ the ion mass, charge and number density, respectively;
time is measured in units of $\omega_c^{-1}=mc/eB_0$, where $B_0$ is the modulus of the background magnetic field ${\bf B}_0$.
Velocities are normalized to the Alfv\'en speed $v_A=B/\sqrt{4\pi m n}=c\omega_c/\omega_p$, and the shock strength is defined by the Alfv\'enic Mach number $M_A=v_{sh}/v_A$, where ${\bf v}_{sh}= -v_{sh}{\bf x}$ is the shock velocity;
we also introduce the energy scale $E_{sh}=\frac m2 v_{sh}^2$.
Ions are initialized with a thermal distribution characterized by a thermal velocity  $\sim v_{A}$, so that the sonic Mach number $M_s$ is roughly equal to $M_A$;
electrons are initially in thermal equilibrium with ions (see Paper I for more details).
In this paper we indicate the shock strength simply with $M=M_A\simeq M_s$.
The shock is generated by the interaction of the primary plasma flow (along $-{\bf x}$) and the counter-streaming flow produced by the reflecting wall set at $x=0$.
The shock propagates to the right in the figures (see Paper I for more details). 

We use very large computational boxes, in order to properly account for the diffusion length of the highest energy ions, and study the time evolution of strong shocks up to $M=100$.
A list of the parameters (Mach number, box size, final time and time step in physical units) of the runs described in the paper is in Table \ref{tab:box}.
High-$M$ shocks are computationally challenging, even for modern supercomputers, since the timestep required to properly conserve energy in hybrid simulations is inversely proportional to the typical velocity of the particles;
therefore, it is necessary to find a trade-off between the box size, in both the longitudinal and transverse dimensions, the shock strength, and the physical time covered by the simulation.
To familiarize the reader with the shock structure that we will be discussing, in this section we present the typical results for a long-term evolution of $M=20$ parallel shocks (i.e., with ${\bf B}_0\parallel {\bf v}_{sh}$).
Then, we investigate the properties of a much stronger shock ($M=100$), in order to show how sensitively the shock dynamics depends on $M$.

\subsection{Long-term evolution of strong shocks}
\begin{figure*}\centering
\includegraphics[trim=0px 0px 0px 0px, clip=true, width=\textwidth]{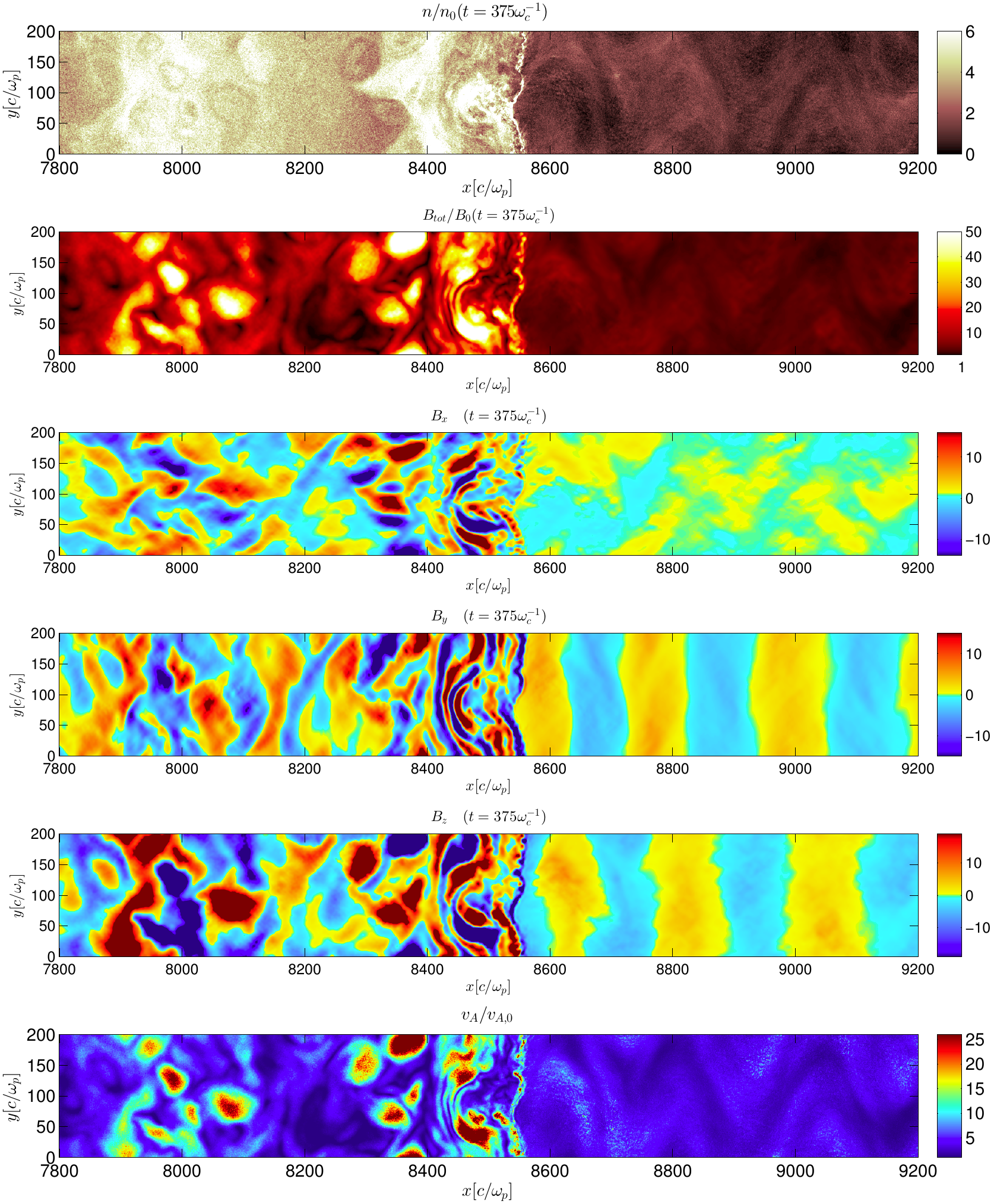}
\caption{\label{fig:summary80}
Relevant physical quantities (as in figure \ref{fig:summary}) for a parallel shock with $M=80$ at $t=375\omega_c^{-1}$ (Run D in Table \ref{tab:box}). 
 The apparent regularity of upstream structures (with respect to figures \ref{fig:summary} and \ref{fig:summary100}) is due to the reduced	 transverse size of the simulation box.}
\emph{A color figure is available in the online journal.}
\end{figure*}

Let us consider a parallel shock with $M=20$, in two different setups: a box of size $(L_x,L_y)=(5\times 10^4,10^3)[c/\omega_p]^2$, in which the shock evolution is followed until $t=1000\omega_c^{-1}$ (hereafter, Run A), and a box of size $(L_x,L_y)=(10^5,10^2)[c/\omega_p]^2$, where the evolution is followed until $t=2500\omega_c^{-1}$ (Table \ref{tab:box}, Run B).
The time step is $\Delta t=5\times 10^{-4}\omega_c^{-1}$ in both cases.
Run A allows us to study the shock dynamics in a box with very large transverse size, fully accounting for the effects of the filamentation instability (CS13), while box B allows us to follow the shock for very long time and to study the development of the non-thermal tail to larger  energies.
We discuss analogies and differences between computational boxes with different transverse sizes in Appendix A.

Figure \ref{fig:summary} shows several physical quantities calculated in Run A (density, magnetic field, Alfv\'en velocity), illustrating the typical magnetic and hydrodynamical structure of an evolved strong shock (at $t=1000\omega_c^{-1}$).
We notice the expected density jump $\sim 4$ at the shock ($x_{sh}\approx 6000 c/\omega_p$), and the distinctive signatures of the filamentation instability induced by accelerated particles streaming ahead of the shock: the cavitation of the upstream, the shock corrugation (which triggers the Richtmeyer--Meshkov instability), and the formation of turbulent structures in the downstream (CS13).

The long-term evolution (up to $t=2000\omega_c^{-1}$) of the post-shock ion spectrum  is shown in figure \ref{fig:evo} (from Run B).
The CR spectrum clearly shows the thermal Maxwellian peak ($p\lesssim mv_{sh}$), and a non-thermal power-law tail, the extent of which grows with time. 
Such a power-law distribution is $f(p)\propto p^{-4}$ in momentum, which corresponds to $f(E)\propto E^{-1.5}$ in energy for non-relativistic particles, and agrees perfectly with the DSA prediction at strong shocks, over more than two decades in energy.
An extensive discussion of the spectrum of the accelerated particles can be found in Paper I.  

\subsection{The high-Mach-number regime}
SNR shocks may have Mach numbers as large as few hundred to a thousand, i.e., they are significantly stronger than $M=20$ shocks discussed above.
As pointed out in Paper I, the acceleration efficiency inferred from simulations is always about 10--15\% for $M\gtrsim 10$.
Conversely, magnetic field amplification is found to be more effective for larger Mach numbers.
We are interested in probing the high-$M$ regime for testing how effective magnetic field amplification can be in SNRs.

Figure \ref{fig:summary100} shows the relevant physical quantities for a very strong parallel shock with $M=100$ (Run C).
The density map (top panel) shows that the asymptotic compression $r\approx 4$ is reached at $x\lesssim 5000 c/\omega_p$. 
The formation of unmagnetized (high $M_A$) shocks is known to be mediated by Weibel instability \citep[e.g.,][]{KT10}.
However, once the shock is well-developed, the likely presence of accelerated ions is expected to amplify the magnetic field in the upstream, eventually affecting the very nature of the shock transition itself. 
We argue that the dramatic filamentation of the upstream produced by accelerated ions (notice the prominent cavities and filaments for $x\gtrsim 8000c/\omega_p$ in figure \ref{fig:summary100}) is a general feature of high Mach number shocks.
Both the thermal plasma and the magnetic field are pushed out of the cavities and accumulated in dense filaments, where the magnetic field can be $\gtrsim 20 B_0$ (second panel of figure \ref{fig:summary100}). 
Even when averaged over the transverse direction, the total magnetic field is more than 5--10 times larger than the initial one, and the region in which $|{\bf B}|\equiv B_{tot}>B_0$ is significantly extended ahead of the shock.
The region at $5000 c/\omega_p\lesssim x \lesssim 8000 c/\omega_p$ represents an extreme case of CR-induced precursor, where the energy in energetic particles is so disproportionately large with respect to the thermal and magnetic components that a dramatic modification of the shock hydrodynamics must occur.
This might also be the manifestation of the \emph{acoustic instability} in a CR precursor, where sound waves may become unstable and form weak ``shocklets'' that significantly heat the upstream plasma \citep{drury-falle86}.

Even if the shock modification induced by CRs is prominent, in our simulations we do not expect CR spectra to become visibly concave as predicted by the non-linear DSA theory \citep[see, e.g.,][]{malkov-drury01}.
Since also accelerated particles are non-relativistic, the adiabatic index of the (gas+CRs) fluid is still 5/3, and the total and subshock compression ratios deviate from $r=4$ by small amounts (because of the modification of the shock jump conditions, see section 6.2 in Paper I);
the corresponding deviation of the spectrum from a power-law is hardly noticeable over (at most) two energy decades.

The $M=100$ case illustrated here serves as a paradigm for very strong shocks investigated in large boxes; however it cannot be followed for very long time due to computational expense.
As a trade-off, we follow the longer-term evolution of a parallel shock with $M=80$ in a box with smaller transverse size (Run D in Table \ref{tab:box}, see figure \ref{fig:summary80}).
The box is large enough to resolve the gyroradius of downstream thermal ions, but not to fully account for the  strong filamentation in the upstream.
Yet, such a run does capture the main features of the regime in which magnetic field amplification is very effective in the precursor ($B_{tot}/B_0\approx 5-10$ on average, with peaks of 15--20$B_0$), and allows us to follow the shock evolution up to $500\omega_c^{-1}$. 
The longer-term evolution of parallel shocks with very high $M$ can eventually be inferred by comparison with the results obtained at lower $M$.

\section{Magnetic field amplification}\label{sec:MFA}

One of the most important problems in particle acceleration at shocks is understanding how effectively CR-induced instabilities amplify the initial magnetic field.
In this section we investigate magnetic field amplification as a function of the shock strength, in a wide range of $M$ up to 100. All the shocks are parallel, and followed until $t=200\omega_c^{-1}$.
The cases $M=100$ and $M=80$ correspond to Run C and D, respectively, while cases with $M=10,20,30,50$ correspond to Run E in Table \ref{tab:box}.

\begin{figure}\centering
\includegraphics[trim=0px 0px 0px 0px, clip=true, width=.5\textwidth]{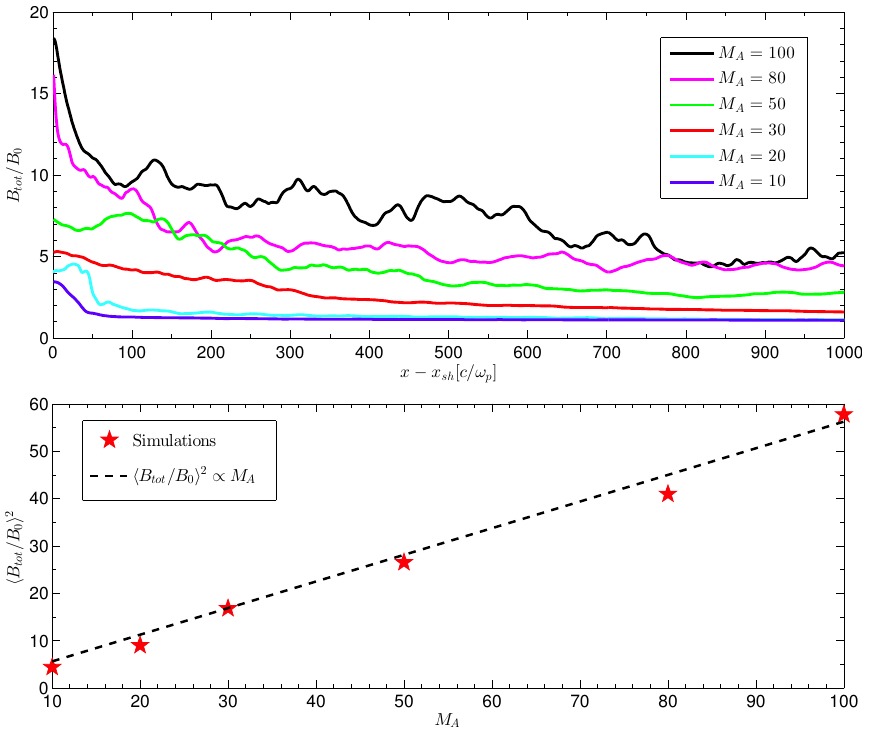}
\caption{\label{fig:dB}
\emph{Top panel}: Magnetic field upstream of the shock at $t=200\omega_c^{-1}$, for different Mach numbers as in the legend. 
$B_{tot}(x)$ is averaged over $200c/\omega_p$ in the transverse size and over $20\omega_c^{-1}$ in time, in order to smooth time and space fluctuations.
\emph{Bottom panel}: Total amplification factor, averaged over a distance $\Delta x=10M_A c/\omega_p$ ahead of the shock, as a function of the Alfv\'enic Mach number (red symbols).
 The dashed line corresponds to $\langle B_{tot}/B_0\rangle^2=0.45 M_A$, and represents the prediction of resonant streaming instability (see eq.~\ref{eq:deltaB}, with $\zeta_{cr}=0.15$).  
\emph{A color figure is available in the online journal}.}
\end{figure}

The top panel of figure \ref{fig:dB} shows the pre-shock profile of the total magnetic field, averaged over $200c/\omega_p$ in the transverse direction, and between $180$ and $200\omega_c^{-1}$ in time.
The position is given as measured from the shock ($x=x_{sh}$), which is estimated by following the peak of the magnetic field intensity (also correlated with the peak in the ion density), and averaging over several tens of $\omega_c^{-1}$.
Tracing the shock position by looking at the maximum gradient in the velocity profile returns very similar results.
Time and space averages are needed to remove the fluctuations induced by filamentary inhomogeneities. 
Even the position of the shock itself may appreciably vary along $y$ in simulations with large transverse size (see, e.g., figure \ref{fig:summary100}), but the profile of the upstream fluid is not very different in any horizontal slice sampling both cavities and filaments.
The most important result is that the magnetic field amplification in the shock precursor is larger for shocks with larger $M$.
The bottom panel of figure \ref{fig:dB} shows the mean value of $B_{tot}/B_0$ over a distance $\Delta x=10 M_Ac/\omega_p$ ahead of the shock, as a function of $M_A$;
we chose integration intervals proportional to the Mach number since the precursor length-scale is larger for stronger shocks.

It is interesting to compare the results in figure \ref{fig:dB} with the prediction of \emph{resonant streaming instability} \citep[e.g.,][]{skilling75a,bell78a,achterberg83}.
By solving the time-independent transport equation for Alfv\'enic modes generated via resonant streaming instability for a shock weakly modified by the CR presence \citep[see, e.g.,][as well as section \ref{sec:res}]{lagage-cesarsky83a,ab06}, 
one gets:
\begin{equation}\label{eq:Pw}
P_w(x)\simeq \frac{P_{cr}(x)}{\tilde{M}_A};\quad P_w(x)=\frac{B_{\perp}^2}{8\pi}
\end{equation}
where $P_w$ and $P_{cr}$ are the magnetic and CR pressures, and $\tilde{M}_A=(1+1/r)M_A$ is the Alfv\'enic Mach number in the shock reference frame (since $r\approx 4$ for strong shocks, typically $\tilde{M}_A\simeq 1.25 M_A$).
We have also introduced the transverse (self-generated) component of the field, whose modulus is $B_{\perp}=\sqrt{B_y^2+B_z^2}$.
If {\bf B} is almost isotropic, one has $B_{\perp}^2\approx\frac23 B_{tot}^2$, and in turn $P_w\approx B_{tot}^2/(12\pi)$.
Normalizing the pressures in eq.~\ref{eq:Pw} to $\rho \tilde{u}^2$, where $\tilde{u}$ is the fluid velocity in the shock frame, and introducing the CR pressure at the shock, $\zeta_{cr}\equiv\frac{P_{cr}(x_{sh})}{\rho \tilde{u}^2}$, one finally obtains:
\begin{equation}\label{eq:deltaB}
\left\langle \frac{B_{tot}}{B_0}\right\rangle_{sh}^2\approx 3 \zeta_{cr} \tilde{M}_A.
\end{equation}
The actual value of $\zeta_{cr}$ can be derived by measuring the deceleration of the fluid in the precursor, and it is strictly related to the CR acceleration efficiency.
In the range of Mach numbers considered here, it varies between 10\% and 15\% at $t=200\omega_c^{-1}$ (see figure 3 in Paper I).
Quite remarkably, plugging $\zeta_{cr}=0.15$ in eq.~\ref{eq:deltaB} provides a very good fit to the amplification factors inferred from simulations (dashed line in figure \ref{fig:dB}).
The extrapolation of eq.~\ref{eq:deltaB} to higher Mach numbers is consistent with the hypothesis that CR-induced instabilities can account for the effective magnetic field amplification inferred at the blast waves of young SNRs, if CR acceleration is efficient.
In particular, a shock velocity of $v_{sh}\approx 4000{\rm kms}^{-1}$ in a medium with $B_0=3\mu$G and $n=1$cm$^{-3}$ corresponds to $M_A\approx600$, and would return $B_{tot}/B_0\approx 20$ for $\zeta_{cr}=0.2$.
It is worth noting that high Mach number shocks show strong filamentary structures (see figure \ref{fig:summary100}), so that physical conditions may vary significantly along $y$;
nevertheless, the present analysis is still expected to hold locally.

\section{Turbulence spectrum}\label{sec:turb}

In this section we investigate the spectrum of the magnetic turbulence generated in the shock precursor by particles accelerated via DSA, for shocks with moderately-large and very large Mach numbers, which show different levels of magnetic field amplification.

Let us start by considering a parallel shock with $M=20$ (Run B), and in particular the self-generated magnetic field ${\bf B}_{\perp}(x)$.
Its spectral energy distribution can be expressed by calculating the Fourier transform of $B_{\perp}(x)$ in the wavenumber  $k$ space\footnote{The spectral energy distribution in $B_{\perp}$ is calculated by summing the spectral energy distribution in $B_y$ and in $B_z$, and does not correspond to the Fourier transform of $B_{\perp}(x)$. 
If $\tilde{B}_{i}(k)$ is the Fourier transform of $B_{i}(x)$, then $\mathcal{F}(k)/k=|\tilde{B}_y(k)|^2+|\tilde{B}_z(k)|^2$.}  
and by posing 
\begin{equation}\label{eq:F}
\frac{B_{\perp}^2}{8\pi}=\frac{B_{0}^2}{8\pi}\int_{k_{min}}^{k_{max}}\frac{dk}{k}\mathcal{F}(k).
\end{equation}
Here $\mathcal{F}(k)$ represents the magnetic energy density per unit logarithmic bandwidth of waves with wavenumber $k$, normalized to the initial energy density $B_0^2/(8\pi)$. 
The maximum wavenumber $k_{max}$ depends on the cell size, and $k_{min}$ on the integration interval.

\begin{figure}\centering
\includegraphics[trim=0px 40px 15px 15px, clip=true, width=.49\textwidth]{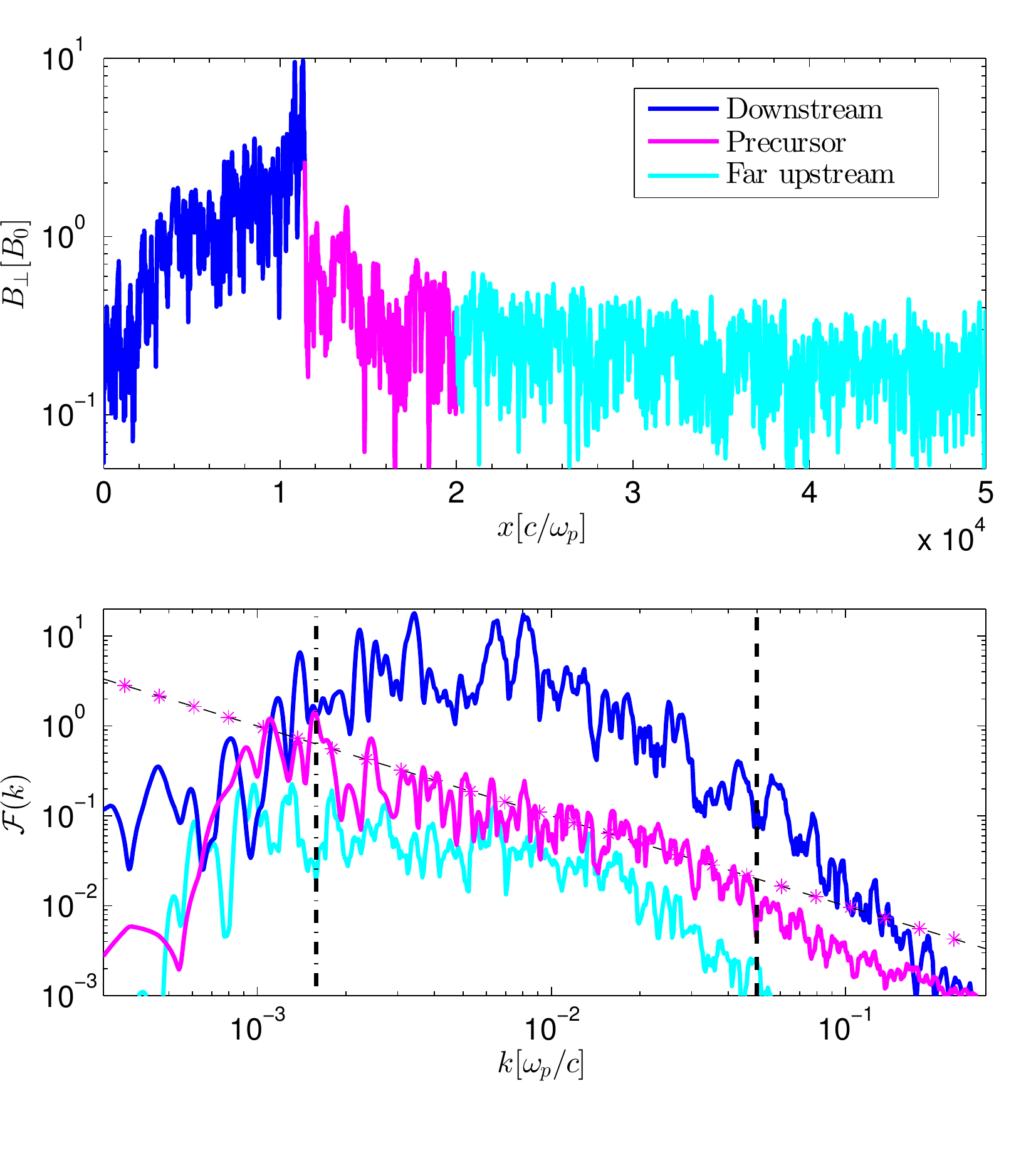}
\caption{\label{fig:Fourier20}
\emph{Top panel}: transverse (self-generated) component of $\bf{B}$ for a $M=20$ parallel shock at $t=2000\omega_c^{-1}$ (Run B). 
\emph{Bottom panel}: power-spectrum of $B_{\perp}$ as a function of wavenumber $k$.
The color code matches corresponding shock regions.
The vertical dashed and dot-dashed line indicate modes resonant with ions of energy $E_{sh}$ and $E_{max}\sim 300E_{sh}$, respectively.
Symbols correspond to $\mathcal{F}(k)\propto k^{-1}$, i.e., the spectral energy distribution produced by  a $\propto p^{-4}$ CR distribution via resonant streaming instability.
\emph{A color figure is available in the online journal.}}
\end{figure}

The top panel of figure \ref{fig:Fourier20} shows the spatial profile of $B_{\perp}(x)$ at $t=2000\omega_c^{-1}$, with the shock at $x_{sh}\sim 10^4 c/\omega_p$.
$\mathcal{F}(k)$ {\bf is calculated at the same $t=2000\omega_c^{-1}$, in three different regions}: the downstream ($0\leq x\leq x_{sh}$), the CR precursor ($x_{sh}\lesssim x\lesssim 2\times 10^4 c/\omega_p$), and the far upstream ($2\times 10^4 c/\omega_p\lesssim x\lesssim 10^5 c/\omega_p$).
The corresponding $\mathcal{F}(k)$ are shown in the bottom panel, with matching color code.
The CR precursor region is chosen in order to encompass the diffusion length of ions with the highest energy, which is of order of $5000c/\omega_p$, as we will discuss in section \ref{sec:NRH}.

First, we focus  on the magenta curve corresponding to the CR precursor.
The wave spectrum is $\mathcal{F}(k)\propto k^{-1}$ (trend defined by the line with symbols in the bottom panel of figure \ref{fig:Fourier20}) between the two vertical lines indicating modes resonant with ions with $E=E_{sh}$ (dashed) and with $E=E_{max}\sim 300 E_{sh}$ (dot-dashed).
We adopt a loose definition of resonance between particles with energy $E$ and modes with wavenumber $k$, namely $kr_L(E,B_0)\sim 1$, which ignores that the local field may be different from $B_0$, and that only the component of ${\bf p}\parallel {\bf B}$ matters for resonant interaction.
$\mathcal{F}(k)$ deviates from the $\propto k^{-1}$ trend for $k\gtrsim 1/r_L(E_{sh})$ and for $k\lesssim 1/r_L(E_{max})$, because of the lack of resonant ions in the precursor.

Besides the normalization, which is directly related to the different magnetic field strength in the precursor and behind the shock, $\mathcal{F}(k)$ has a similar shape throughout the simulation box (see different curves in figure \ref{fig:Fourier20}).
The far upstream cyan curve shows a high-$k$ steepening at a wavenumber resonant with ions with $\sim 10 E_{sh}$ rather than with $E_{sh}$, consistent with the fact that low-energy CRs do not make it far upstream. 
 
\begin{figure}\centering
\includegraphics[trim=0px 40px 15px 15px, clip=true, width=.49\textwidth]{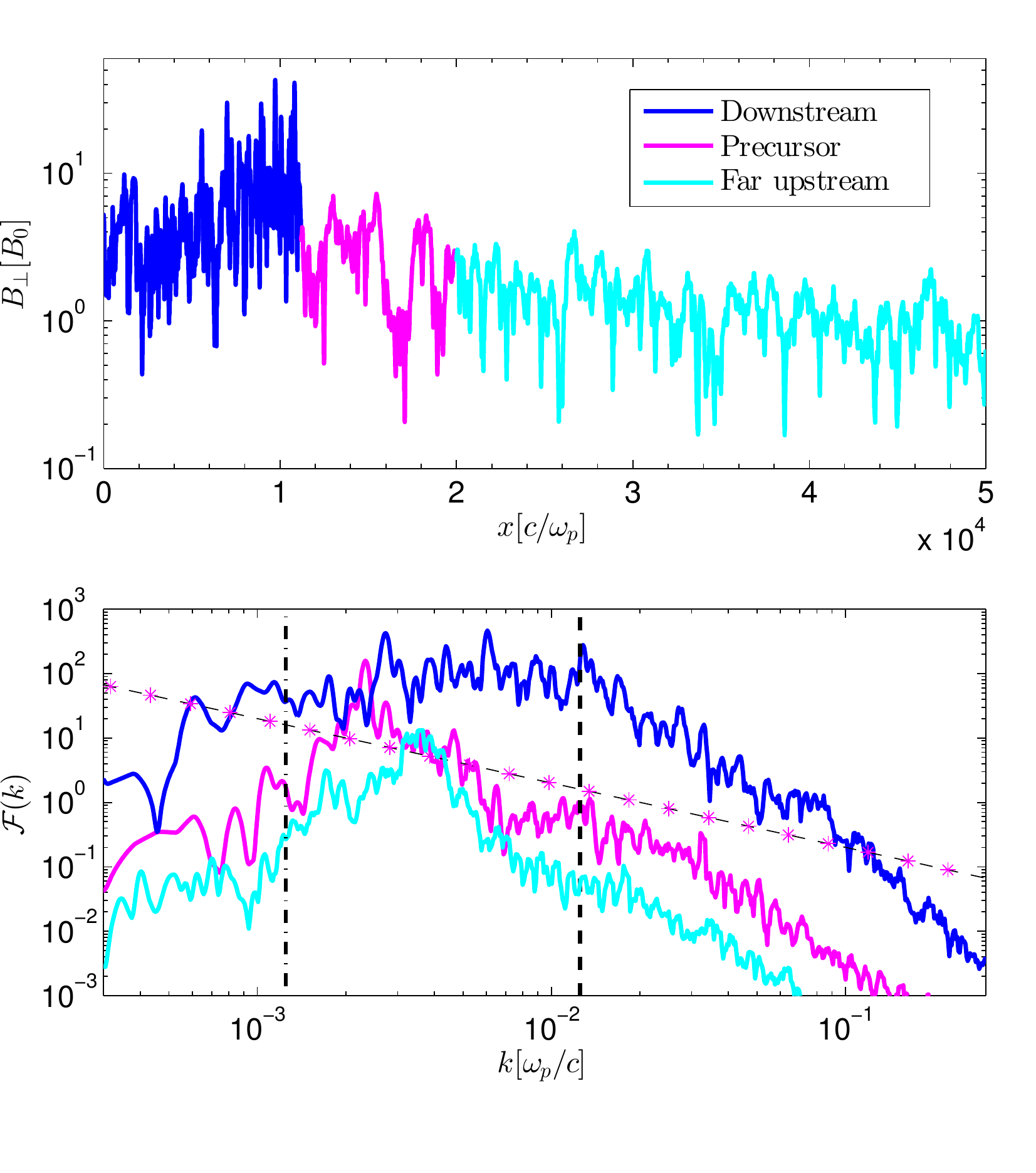}
\caption{\label{fig:Fourier80}
As in figure \ref{fig:Fourier20} for a parallel shock with $M=80$, at $t=500\omega_c^{-1}$, when $E_{max}\approx 100E_{sh}$ (Run D). 
The magnetic field is significantly more amplified than in the $M=20$ case, with $\mathcal{F}(k)$ in the precursor a factor of about 10 larger than in  figure \ref{fig:Fourier20}.
Note that the resonance at $E_{max}$ (dot-dashed line) is calculated in $B_0$: accounting for the amplified field would shift the resonance at higher $k$.
\emph{A color figure is available in the online journal.}}
\end{figure}

Let us consider now the case of a stronger parallel shock with $M=80$ (Run D), where magnetic field amplification is more efficient.
The power spectrum $\mathcal{F}(k)$ at $t=500\omega_c^{-1}$ is shown in figure \ref{fig:Fourier80}.
The biggest difference with respect to the $M=20$ case is that $\mathcal{F}(k)$ is more than a factor of 10 larger in the precursor (magenta curve).
Since $\mathcal{F}\propto (B_{tot}/B_0)^2$, such a result is consistent with the measurements of magnetic amplification in section \ref{sec:MFA} (see figure \ref{fig:dB}).
Most of the energy in magnetic turbulence is still at wavenumbers resonant with accelerated particles (between the vertical lines); however, the peak in $\mathcal{F}(k)$ is not exactly at $kr_L(E_{max},B_0)\sim 1$, but at a slightly higher $k$.
This is just the effect of the actual field in the precursor being few times $B_0$, and is consistent with most of the energy being in waves resonant with highest-energy ions, as for $M=20$.
Instead, the peak in the wave spectrum in the far upstream (cyan curve in figure \ref{fig:Fourier80}) is at wavenumbers a factor of 2--3 larger than in the precursor (magenta curve): 
such an effect cannot be ascribed to the local magnetic field being much larger than $B_0$, but rather contains information about the nature and the evolution of unstable modes in the far upstream, as we comment in section \ref{sec:NRH}.  

\subsection{\label{sec:res} Resonant streaming instability}
In order to understand how magnetic energy is distributed in wavelength, we consider the stationary equation for the growth and transport of magnetic turbulence in the upstream fluid \cite[see, e.g.,][]{mckenzie-volk82}:
\begin{equation}\label{transw0k}
	\frac{\partial \varepsilon(k,x)}{\partial x}=
	u(x)\frac{\partial \mathcal F(k,x)}{\partial x}+
	\sigma(k,x) \mathcal F(k,x)\;,
\end{equation}
where $\varepsilon(k,x)$ and $\mathcal F(k,x)$ are the energy flux and pressure per unit logarithmic bandwidth of waves with wavenumber $k$, and $\sigma(k,x)$ is the rate at which the energy in magnetic turbulence grows;
In Equation \ref{transw0k} we have not explicitly included the damping of magnetic modes, which is inferred to heat the precursor up, keeping magnetic and gas pressure in equipartition (see \S6.1 in Paper I);
therefore, our formulas describing the level of magnetization inferred in simulations effectively include wave damping.
The growth-rate of Alfv\'en waves produced by resonant streaming instability reads \citep[e.g.,][]{skilling75b,bell78a,achterberg83}:
\begin{equation}\label{eq:sigmak}
	\sigma(k,x)=\frac{4\pi}{3}\frac{v_A}{P_{w,0}\mathcal F(k,x) }
	\left[p^4v(p)\frac{\partial f(x,p)}{\partial x}\right]_{p=\bar{p}_k}\,,
\end{equation}
where $P_{w,0}=B_0^2/(8\pi)$, $f(x,p)$ is the isotropic part of the local ion distribution, and $\bar{p}_k=m\omega_c/k$ is the resonance condition. 
Quantities are measured in the shock frame, where stationarity is achieved \citep[see][for the solution of eq.~\ref{eq:sigmak} in the presence of efficient CR acceleration]{jumpkin}. 
Assuming equipartition between electromagnetic and kinetic energy density in the waves, we have $\varepsilon(k,x)\approx 2u \mathcal{F}(k,x)$ and we can rewrite eq.~\ref{transw0k} as 
\begin{equation}\label{eq:Fkdiff}
	u(x)P_{w,0}\frac{\partial \mathcal F(k,x)}{\partial x}=v_A\frac{\partial \mathcal P (\bar{p}_k,x)}{\partial x}\,,
	\end{equation}
where $\mathcal P(x,p)$ expresses the pressure in CRs per unit logarithmic momentum bandwidth, i.e.,	
\begin{equation}	
	P_{cr}(x)=\frac{1}{3} \int dp 4\pi p^2 v(p)p f(x,p)\equiv\int \frac{dp}{p} \mathcal P(p,x)\,.
\end{equation}
Neglecting the shock modification in the precursor (i.e., taking constant fluid and Alfv\'en velocity) and assuming that both $\mathcal P$ and $\mathcal F$ vanish at upstream infinity,  integration of eq.~\ref{eq:Fkdiff} is straightforward and returns 
\begin{equation}\label{eq:Fk}
P_{w,0}\mathcal F(k,x)=\frac{v_A}{u}\mathcal P(\bar{p}_k,x)\,.
\end{equation}
Finally, integrating eq.~\ref{eq:Fk} over resonant $k$ and $p$ gives eq.~\ref{eq:Pw}.
Eq.~\ref{eq:Fk} states an important fact: \emph{the spectral energy density in magnetic turbulence excited via resonant streaming instability is proportional to the energy density in CRs at the corresponding resonant momenta}.
For a $f(p)\propto p^{-4}$ spectrum of non-relativistic ($v=p/m$) particles, $\mathcal P(p)\propto p$, and most of the energy is at the highest momenta; 
therefore, the corresponding wave spectrum is expected to be $\mathcal F(k)\propto k^{-1}$ in the shock precursor\footnote{Note that, in the relativistic regime where $v\simeq c$, a CR spectrum $f(p)\propto p^{-4}$ would correspond to constant energy per momentum decade, and in turn to a $\mathcal F(k)$ flat in wavenumber.}.
The agreement between such a $\propto k^{-1}$ trend and the wave spectrum in the CR precursor for our $M=20$ run is remarkable (compare the magenta curve with the symbols in the bottom panel of figure \ref{fig:Fourier20}).
The scaling is less evident for the $M=80$ case (figure \ref{fig:Fourier80}) where the CR spectrum is steeper than $p^{-4}$ because the non-thermal tail is not fully developed yet (see Paper I).

The good agreement of simulations with the scaling $\delta B/B_0\propto \sqrt{M_A}$ (eq.~\ref{eq:deltaB}) suggests that some form of resonant streaming instability should be prominent in CR precursors of SNR shocks.
However, it is not entirely obvious whether the Alfv\'en velocity in eq.~\ref{eq:Fk} should be calculated with $B_0$ even in the nonlinear regime.
PIC simulations in periodic boxes showed that the Alfv\'en velocity grows proportionally to the magnetic field  in the nonlinear stage of the instability \citep{rs09}, and the enhanced phase velocity of self-generated modes may play an important role in explaining the steep ion spectra observed in $\gamma$-ray bright SNRs \citep{gamma, efficiency}.
Simulations presented here are not conclusive in this respect: longer runs of strong shocks are needed to convincingly claim a (possible) steepening of about $10-20\%$ in the CR spectral slope with respect to the canonical value of 4.

Finally, we point out that the magnetic spectrum has non-negligible power even at scales $\gtrsim r_L(E_{max})$: in figure \ref{fig:Fourier80}, $\mathcal F(k)\gtrsim 0.1$ for $1/r_L(100E_{max})\lesssim k\lesssim 1/r_L(E_{max})$.
These modes may either be driven  by escaping ions with energy larger than $E_{max}$, or be the signature of a large-wavelength instability, like the \emph{firehose} instability,  \citep[see, e.g.,][]{blandford-eichler87,shapiro+98}.

\subsection{Dependence on the shock obliquity}

As shown in Paper I, the amount of magnetic turbulence triggered by CR instabilities strongly depends on the shock obliquity, being mostly prominent for parallel and quasi-parallel shocks.
At quasi-perpendicular shocks, instead, few or no accelerated particles propagate into the upstream and the field is not amplified.

\begin{figure}\centering
\includegraphics[trim=0px 40px 20px 260px, clip=true, width=\columnwidth]{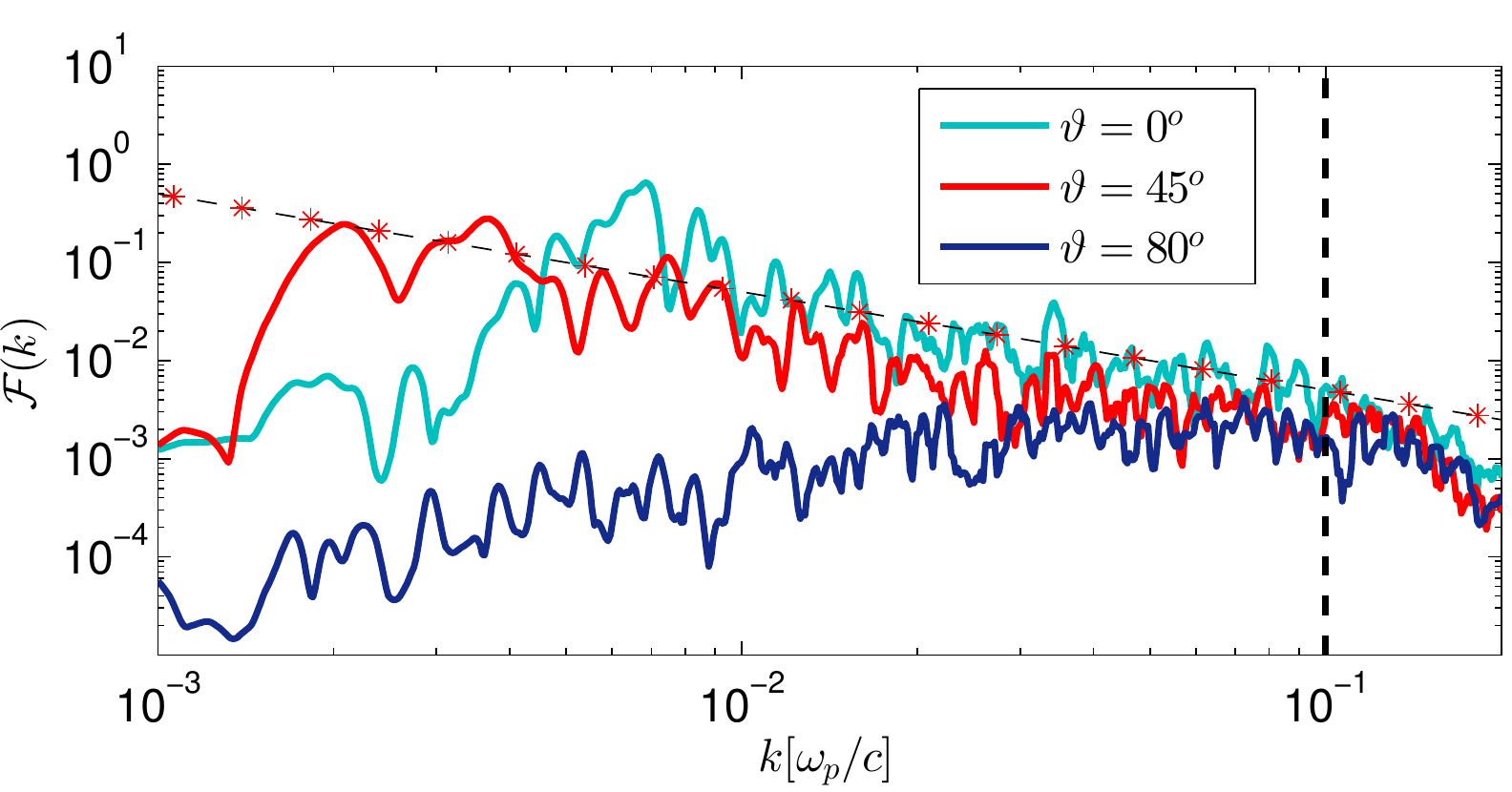}
\caption{\label{fig:Fourier_th}
Power spectrum of the self-generated $\textbf{B}$, calculated in a region of $5000c/\omega_p$ upstream of the shock ($M=10$, run E), for three different obliquities as in the legend.
The dashed line corresponds to wavenumber $k\approx 1/r_L(E_{sh})$, and symbols correspond to $F(k)\propto k^{-1}$.
Configurations at $\vartheta=0\deg$ and 45$\deg$ are quite similar to each other and agree nicely with DSA prediction, while the magnetic energy for the quasi-perpendicular shock is significantly smaller.
\emph{A color figure is available in the online journal.}}
\end{figure}

Figure \ref{fig:Fourier_th} shows the spectrum of the self-generated magnetic turbulence (with $B_{\perp}=B_z$), calculated  in a region of thickness $5000c/\omega_p$ upstream of the shock at $t=200\omega_c^{-1}$, for different shock inclinations $\vartheta=0\deg,45\deg,80\deg$; here, $\vartheta $ is the angle between ${\bf v}_{sh}$ and ${\bf B}_0=(B_0\cos{\vartheta},B_0\sin{\vartheta},0)$.
All of the runs have $M=10$, box size $(L_x,L_y)=(40000c/\omega_p, 500c/\omega_p)$ and time-step $\Delta t=10^{-3} \omega_c^{-1}$ (Run E in Table \ref{tab:box}).
As in figure \ref{fig:Fourier20}, the vertical dashed line marks the wavenumber resonant with ions with energy $E_{sh}$.
The $\vartheta=0\deg$ and $\vartheta=45\deg$ cases are similar to each other, and agree nicely with the $\mathcal F(k)\propto k^{-1}$ trend discussed above.
This is not surprising, since at $\vartheta=45\deg$ the shock is still quite efficient in accelerating particles, even if filamentation instability is moderately suppressed and $\delta B/B_0$ is smaller than in the parallel case (see Paper I).
The inclined geometry of the background field facilitates the return of high-energy particles, actually halving the diffusion time for $\vartheta=45\deg$, and allowing the achievement of an $E_{max}$ twice as large as in the parallel case.
This effect is reflected also in the wave spectrum in figure \ref{fig:Fourier_th}, with the $\vartheta=45\deg$ curve peaking at a  wavenumber lower than for $\vartheta=0\deg$.
For $\vartheta=80\deg$ (blue curve in figure \ref{fig:Fourier_th}), instead, there are no accelerated ions streaming ahead of the shock, and no magnetic turbulence is effectively generated $(\mathcal{F}(k)\lesssim 10^{-3}$ for $kr_L(E_{sh})\lesssim 1)$.

We conclude that resonant modes are excited for all shock inclinations $\vartheta\lesssim 45\deg$, where DSA acceleration is efficient. 
On one hand, at parallel shocks filamentation instability is more effective and eventually leads to larger amplification factors (CS13); 
on the other hand, the field geometry of more oblique shocks prevents CRs from diffusing far from the shock, potentially reducing their acceleration time by a factor of $\sim \cos\vartheta$.
Above $\vartheta\sim 45\deg$, however, DSA is ineffective.
As shown in Paper I, ions gain a factor of a few in energy because of shock drift acceleration, but their current does not perturb the large-scale magnetic configuration.

\section{The role of NRH modes}\label{sec:NRH}
Both the level of amplification (eq.~\ref{eq:deltaB}) and the spectrum of the self-generated magnetic turbulence in the precursor (figures \ref{fig:Fourier20} and \ref{fig:Fourier80}) are consistent with resonant streaming instability being the channel through which CRs amplify the initial magnetic field.
However, there is a short-wavelength instability \citep[the \emph{non-resonant hybrid}, NRH, instability,][]{bell04,bell05} that is predicted to grow faster than the resonant one.

Different flavors of streaming instability can be obtained by linearizing the dispersion relations of circularly polarized waves with wave vectors aligned with a background magnetic field \cite[see, e.g.,][]{krall-trivel}, also including the return current of electrons balancing the positive CR current \citep[see, e.g.,][for a detailed kinetic derivation]{ab09}.
In this framework, the resonant (non-resonant) branch corresponds to waves for which the electric and magnetic field vectors, at a fixed point in space, rotate in the same sense as protons (electrons); such modes are usually referred to as left-handed (right-handed).
In the quasi-linear limit, the maximum growth rate $\Gamma$ contributed by the current of ions with momentum $p$ propagating upstream with velocity $v_{cr}$ reads \citep[see eq.~28 in][]{ab09}\footnote{The numerical factor in $\Gamma_{res}$ should read $\sim\sqrt{0.43}$, here approximated as $1/\sqrt{2}$ for simplicity and analogy with $\Gamma_{nrh}$.}:
\begin{equation}\label{eq:gammas}
\frac{\Gamma_{res}(p)}{\omega_c}\simeq \sqrt{\xic(p)\frac{mv_{cr}}{2p}}, \quad
\frac{\Gamma_{nrh}(p)}{\omega_c}\simeq \frac{\xic(p)}{\sqrt{2}}\frac{v_{cr}}{v_A},
\end{equation}
where we introduced the number density of CRs with momentum larger than $p$ normalized to the density of the background plasma, $\xic(p)\equiv n_{cr}(>p)/n$.
Here $v_{cr}$ is the bulk CR velocity as seen by the waves advected with the upstream fluid: for diffusing CRs that are almost isotropic in the shock reference frame, $v_{cr}\approx v_{sh}$, while for escaping (i.e., free-streaming) CRs, $v_{cr}\approx \sqrt{2E_{max}/m}$ ($v_{cr}\approx c$ for relativistic particles).
The most-unstable wavenumbers corresponding to the growth rates in Equation \label{eq:gammas} are $K_{res}\simeq 1/{r_L(p)}$ and $K_{nrh}\simeq \Gamma_{nrh}/{v_A}$.
 Strictly speaking, the NRH instability at $K_{nrh}$ depends on the current due to all of the streaming CRs, while the resonant instability is driven only by CRs with $p\gtrsim 1/K_{res}$.
While not completely accurate in the general case, $\xic$ provides a good description of the CR content both close to the shock, where $p\sim mv_{sh}$, and far upstream, where $p\sim p_{max}$.

The ratio of the growth rates of NRH and resonant instability in eq.~\ref{eq:gammas} is:
\begin{equation}\label{eq:Wp}
W(p)=\frac{\Gamma_{nrh}}{\Gamma_{res}}\simeq\frac{\sqrt{\xic(p) v_{cr}p/m}}{v_A}
\simeq \m\sqrt{\xic(mv_{cr})},
\end{equation}
where we assumed $\xic(p)\propto 1/p$ as for a $f(p)\propto p^{-4}$ CR distribution, and 
introduced the effective Alfv\'enic Mach number of the CR current, $\m\equiv v_{cr}/v_A$.
For diffusing CRs, which are almost isotropic in the shock frame and move with $v_{cr}\simeq v_{sh}$ with respect to the upstream fluid, $\m\approx M_A$, and $\xic(mv_{sh})\approx 10^{-3}$ (see Paper I).
This means that in precursors of shocks with $M_A\gtrsim W/\sqrt{\xic}\approx 30$ the NRH instability grows faster than the resonant one \citep[see also][]{ab09}.

The crucial questions that we want to address are: what is the maximum level of amplification achievable via NRH instability upstream of a SNR shock? And, what is the relative contribution of resonant and NRH instabilities in different shock regions? 
In order to answer these questions in a quantitative way, we need to know how the growth rate and the maximally-growing mode evolve when the instability enters its nonlinear stage, i.e., when $b\equiv B_{\perp}/B_0\gtrsim 1$.

\cite{rs09} have derived the nonlinear dispersion relation for NRH modes, from which it is possible to work out phase velocity and growth rate ($\omega/k$ and $\gamma$) for arbitrary amplification factor $b$.
From the imaginary part of eq.~A12 in appendix A of their work, one can write and solve a differential equation for $\omega(b)$ for the fastest-growing mode (eq.~2 in the same paper); 
by inserting such a solution for $\omega(b)$ in the real part of equation A12, we eventually find $\gamma(b)$, which in the limit  $v_A\ll c$ reads:
\begin{equation}
\frac{\gamma^2}{v_A^2}=\frac{k(2K_0-k)}{b^2+1}-\frac{K_0^2}{\m^2}\frac{2b^2+1}{(b^2+1)^2};~ K_0\simeq\omega_c\xic \frac{\m_0}{v_{A,0}}
\end{equation}
where the subscript 0 labels initial quantities ($b\ll 1$), $\gamma(k)$ is the growth rate of the mode with wavenumber $k$, and $K_0$ is the fastest-growing mode.
For $b\gtrsim 1$, one has $v_A^2\approx v_{A,0}^2(b^2+1)$, and the growth rate of the fastest-growing mode, $\Gamma(b)= \gamma(K_0,b)$, can be written as:
\begin{equation}\label{eq:Gamma}
\Gamma(b)\simeq v_{A,0}K_0\sqrt{1-\frac{2b^2+1}{\m_0^2}}
\approx\Gamma_0\sqrt{1-\frac{2b^2}{\m_0^2}},
\end{equation}
where the last equations holds for $1\ll 2b^2\leq \m_0^2$.
With eq.~\ref{eq:Gamma} we can calculate the evolution of the amplification factor in the nonlinear regime by integrating $\dot{b}(t)=\Gamma b(t)$, with $t=0$ defined as the time when $b(0)\approx 1$, obtaining
\begin{equation}\label{eq:bt}
b(t)\simeq\frac{e^{\Gamma_0t}[\m_0^2-\m_0\sqrt{\m_0^2-2}]}{1-e^{2\Gamma_0t}[\m_0^2-1+\m_0\sqrt{\m_0^2-2}]}.
\end{equation}
First, we notice that the maximum amplification factor provided by the NRH instability is $b_{max}\simeq\m_0/\sqrt{2}$, which corresponds to $\Gamma(b_{max})\simeq 0$; 
second, such a maximum is reached for $\Gamma_0t_{max}\simeq \log({\sqrt{2}\m_0})$, which gives us the estimate of the duration of the exponential phase.
These results are in good agreement with PIC and hybrid simulations with controlled ion beams \citep[e.g.,][]{rs09,gargate+10}, which show that $b_{max}\approx \m_0$ at saturation.
Simulations also show that the exponential phase lasts for $\Delta t\approx 3-5\Gamma_0^{-1}$, and that the instability saturates after few $t_{max}$. 
The additional growth observed after $t_{max}$ can be understood by accounting for the contribution of modes with $k\lesssim K_0$, whose slower growth rate is $\gamma(k)\simeq \Gamma\sqrt{k/K}$.  

\subsection{The free-escape boundary}
An important difference between controlled simulations in boxes with periodic boundary conditions and realistic shock precursors is that the CR current is not fixed, but is rather determined by scattering in the self-generated turbulence.
The most unstable mode $K_0$ is unable to deflect current ions because it is right-handed and has small-wavelength, namely $K_0r_L(v_{cr})\simeq \xic \m^2\gg 1$.
Nevertheless, PIC simulations show that the most unstable wavenumber decreases as $K(b)\simeq K_0/b^2$ \citep{rs09}, which implies $K(b)r_L(b)\propto b^{-3}$.
There exists a critical amplification factor $b_*\simeq \sqrt[3]{\xic \m_0^2}$ for which the wavelength of NRH modes becomes comparable with the gyroradius of current CRs, i.e., $K(b_*)r_L(b_*)\approx 1$.
In the nonlinear stage, right-handed NRH modes may become \emph{resonant in wavelength} with ions driving the instability; when this happens, such ions are effectively scattered, and the current is disrupted.
Since $\xic \m_0\lesssim 1$ is a necessary condition for the growth of the NRH instability to nonlinear levels \citep[see, e.g., sec.~3.1 of][]{rs09}, one has a limit on $b$ independent of the CR density: $b_*\leq \sqrt[3]{\m_0}$.
For relativistic CRs in the ISM, we have $b_*\lesssim\sqrt[3]{c/v_A}\approx 20-30$, which means that CRs escaping from a SNR might be able to pre-amplify the ISM field by more than one order of magnitude before being isotropized \citep[see also][]{bell+13}. 

In realistic shock precursors there must be two distinct regions: i) the \emph{far upstream} region, where the current is provided by \emph{free-streaming} ions with $E\gtrsim E_{max}$ that excite small-wavelength NRH modes, and ii) the \emph{CR precursor}, where the current is sustained by \emph{diffusing} CRs with $v_{cr}\simeq v_{sh}$.   
The boundary between the two regions is marked by the condition $b\approx b_*$, and represents the \emph{free-escape boundary} widely adopted in DSA theory \citep[see, e.g.,][and references therein]{comparison}.
Moreover, any dynamical modification induced by energetic particles occurs in the precursor, where CRs are well magnetized and can exert pressure on the incoming fluid (see Paper I);
escaping particles can still remove energy from the system, making the shock behaving as partially-radiative \citep[see, e.g.,][]{escape}.

We searched for NRH modes in our long global simulations, which also account for shock evolution and self-consistent CR currents, by considering the different polarization of resonant and NRH modes \citep[see, e.g.,][]{gs12}.
From eq.~\ref{eq:Wp} we expect NRH modes to be prominent only for $M\gtrsim 30$;
in the upstream of $M=20$ shocks (Runs A and B) we find regions where wave polarization is mainly left-handed (i.e., resonant with accelerated ions), while for $M=80$ (Run D) modes are predominantly right-handed, i.e., generated by the NRH instability.
Also, while for shocks with $M=20$  the spectral power density $\mathcal F(k)$ can be explained as due to resonant streaming instability (figure \ref{fig:Fourier20}), this is not the case for higher Mach numbers.
In the far upstream of the $M=80$ case, $\mathcal F(k)$ has a peak at wavelengths smaller than the gyroradius of ions with $E_{max}\sim 100 E_{sh}$ (figure \ref{fig:Fourier80}).
Such a peak moves to longer wavelengths closer to the shock, eventually matching the wavenumber of modes resonant with $E_{max}$.
In Run D, the density of escaping CRs is $\xic(E\geq E_{max})\approx 10^{-4}$, and for them $\m_0\simeq M_A\sqrt{E_{max}/E_{sh}}$;
therefore, the growth rate of the fastest-growing mode is $\Gamma_0\approx 0.07\omega_c$.
Run D is long and large enough for the fastest-growing mode to reach saturation, since time and length scales are larger than $t_{sat}\approx\log(\m)/\Gamma_0\approx 100\omega_c^{-1}$ and $L_{sat}\simeq v_{sh}t_{sat}\approx 8000c/\omega_p$. 
With the same parameters, we estimate $b_*\approx 3.7$, which nicely matches the level of field amplification at the boundary between precursor and far upstream, which in figure \ref{fig:Fourier80} is set at $x\approx 2\times 10^4c/\omega_p$.
 
The determination of the free-escape boundary for shocks with low Mach numbers cannot rely on the migration to longer wavelengths of NRH modes, since field amplification typically proceeds in the linear regime, and resonant and NRH instabilities grow at almost the same rate.
In this case, the current in escaping ions is more easily disrupted, and the shock precursor extends for about one diffusion length of the ions with maximum energy; 
for an extended characterization of particle diffusion, and in particular for the parametrization of the diffusion coefficient inferred from simulations of shocks with low and high Mach numbers, we remand to Paper III.
 
We conclude this section with some caveats.
First, most of the findings above are based on 2D simulations with limited transverse size and up to $M=80$.
The shock in Run C, which has $M=100$ and very extended transverse size (figure \ref{fig:summary100}), shows how filamentation may be important for very strong shocks.
The complex pattern of cavities and filaments suggests that 1D descriptions may not properly capture growth and saturation of the magnetic turbulence.
Nevertheless, filamentation \emph{enhances} the production of magnetic turbulence, both upstream and downstream (CS13); 
we argue that simulations with limited transverse size place lower limits on magnetic field amplification and CR scattering.
Second, one may question the applicability of the approach of section \ref{sec:res}, which only includes resonant instability due to diffusing CRs, to very strong shocks.
Eqs.~\ref{transw0k}--\ref{eq:Fk} should still be appropriate in the precursor, which we defined as the region of the upstream where CRs with $E\lesssim E_{max}$ are effectively scattered.
Moreover, since the free-escape boundary is determined by the condition that NRH modes become comparable in wavelength to CR gyroradius, $K r_L(E_{max})\approx 1$, in the precursor the resonant and NRH instability must grow at a comparable rate.
The proper transport equation for the upstream magnetic turbulence should include also the current in escaping CRs and migration in wavelength, which is indeed crucial for regulating escape and scattering of the most energetic CRs.
However, we argue that eq.~\ref{eq:sigmak} still captures the order of magnitude of the relevant growth rate in the precursor, thereby providing a good fit of the scaling of the total amplification factor at the shock (figure \ref{fig:dB}). We defer to forthcoming works a more detailed description of the interplay among NRH, resonant and  filamentation instabilities, the possible role of long-wavelength (firehose-like) instabilities, and the extension of the present results in 3D simulations.

\section{Conclusions}\label{sec:concl}
This paper is the second of a series aimed to investigate several aspects of ion acceleration at non-relativistic shocks through hybrid simulations. 
The first paper \citep[][Paper I]{DSA} measured the spectrum of the accelerated particles, the dependence of the acceleration efficiency on shock strength and inclination with respect to the upstream magnetic field, and the shock modification induced by efficient CR acceleration.
In this paper we investigate the magnetic turbulence generated by super-Alfv\'enic particles in the shock precursor.
To perform this study in a consistent way, in principle one needs to: 
1) follow the shock for very long time in physical units, which requires a box sufficiently large in the direction of the shock propagation, in order to see the development of the non-thermal ion tail; 
2) use large boxes in the transverse directions, in order to retain the modifications induced by the filamentation instability \citep{filam};
3) run shocks with very high Mach-numbers, in order to simulate conditions relevant to real SNR blast waves.
Since it is computationally impossible to satisfy all of these requirements in the same run, we individually explored these limits in different state-of-the-art simulations, always bearing in mind the physics that may be missing when one or more of the points above is neglected.
Our main findings are the following.
\begin{itemize}
\item High Mach number shocks ($M\gtrsim 50$) can produce very strong CR-induced precursors (e.g., figure \ref{fig:summary100}), in which the incoming plasma is dramatically slowed down and heated up (see Paper I).

\item Magnetic field amplification ahead of the shock is more effective for strong shocks. 
The amplification factor averaged over regions comparable with CR diffusion lengths is $\sim 10$ for $M_A=100$, and  scales as $(B_{tot}/B_0)^2\propto M_A$ (figure \ref{fig:dB}). 
The extrapolation of this trend to the Mach numbers of a few hundred relevant for young SNRs can account for their large inferred magnetic fields of few hundred $\mu$G.

\item Upstream of shocks with $M\lesssim 30$, the spectrum of excited magnetic turbulence (figure \ref{fig:Fourier20}) is consistent with the prediction of resonant streaming instability \citep[e.g.,][]{bell78a,achterberg83};
the energy distribution in waves is determined by the energy distribution in accelerated particles (eq.~\ref{eq:Fk}).
Amplification vanishes for quasi-perpendicular shocks, where acceleration is inefficient and the CR current is weak or even absent.

\item Shocks with $M_A\gtrsim 30$ show a different behavior because small-wavelength modes excited by the non-resonant hybrid (NRH) instability grow faster than resonant ones  \citep{bell04,bell05}.
NRH modes are excited by the escaping ions (which have energies close to the maximum energy $E_{max}$) and their wavelength increases $\propto (\delta B/B_0)^2$ until it becomes comparable with the gyroradius of ions with $E_{max}$.

\item For such high-$M_A$ shocks, we can distinguish two regions: the far upstream region, where the current is provided by escaping CRs, and the precursor, where the current is provided by the gradient in diffusing (magnetized) CRs.
NRH instability dominates far upstream, while in the precursor resonant and NRH instabilities grow at a comparable rate.
The interface between the two regions represents the so-called \emph{free-escape boundary}.
\end{itemize}

These results can be used to include self-consistent microphysics into models of particle acceleration at shocks, especially into non-linear approaches to DSA \citep[see, e.g.,][for a comparison of different techniques]{comparison}, which typically require prescriptions for the position of the free-escape boundary, and for the dominant channels of magnetic field amplification.
The total level of magnetic field amplification, and its scaling with the shock Alfv\'enic Mach number, is also crucial in modeling synchrotron emission of non-thermal electrons accelerated in non-relativistic shocks, for instance in SNRs, AGN lobes, and galaxy clusters.
In forthcoming publications we will cover diffusion of particles in the self-generated magnetic turbulence and the evolution of the maximum CR energy with time \citep{diffusion}, and the mechanisms that lead to the injection of ions into DSA, in order to provide closure for the present series of papers. 


\subsection*{}
We wish to thank L.\ Gargat\'e for providing a version of \emph{dHybrid}, P.\ Blasi, E.\ Amato and A.\ Bell for stimulating discussions, and the referee for the thorough comments and suggestions. 
This research was supported by NSF grant AST-0807381 and NASA grant NNX12AD01G, and facilitated by the Max-Planck/Princeton Center for Plasma Physics. 
This work was also partially supported by a grant from the Simons Foundation (grant \#267233 to AS), and by the NSF under Grant No.\ PHYS-1066293 and the hospitality of the Aspen Center for Physics.
Simulations were performed on the computational resources supported by the PICSciE-OIT TIGRESS High Performance Computing Center and Visualization Laboratory. This research also used the resources of the National Energy Research Scientific Computing Center, which is supported by the Office of Science of the U.S. Department of Energy under Contract No.\ DE-AC02-05CH11231, and XSEDE's Ranger and Stampede under allocation No.\ TG-AST100035.

\appendix
\section{Dependence on the transverse size of the box}
\begin{figure}\centering
\includegraphics[trim=0px 0px 0px 0px, clip=true, width=.48\textwidth]{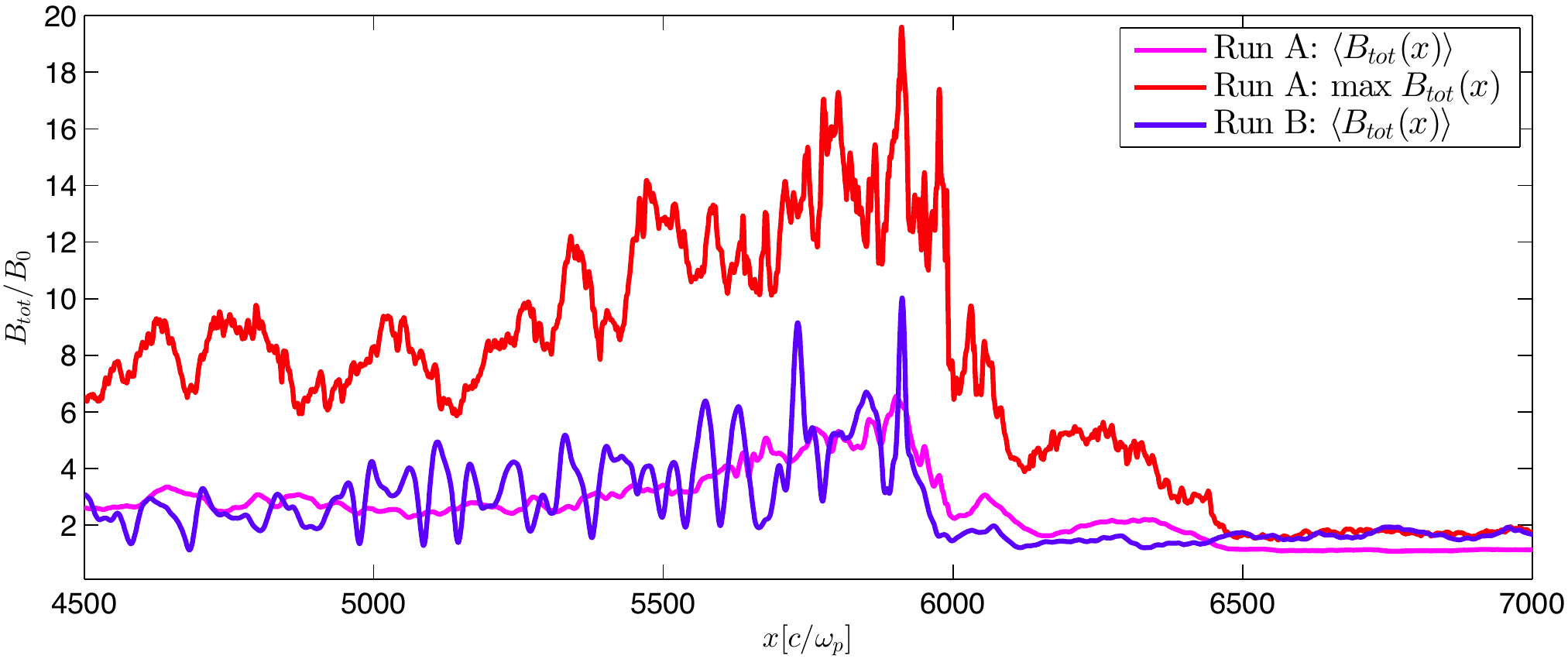}
\includegraphics[trim=0px 3px 0px 0px, clip=true, width=.49\textwidth]{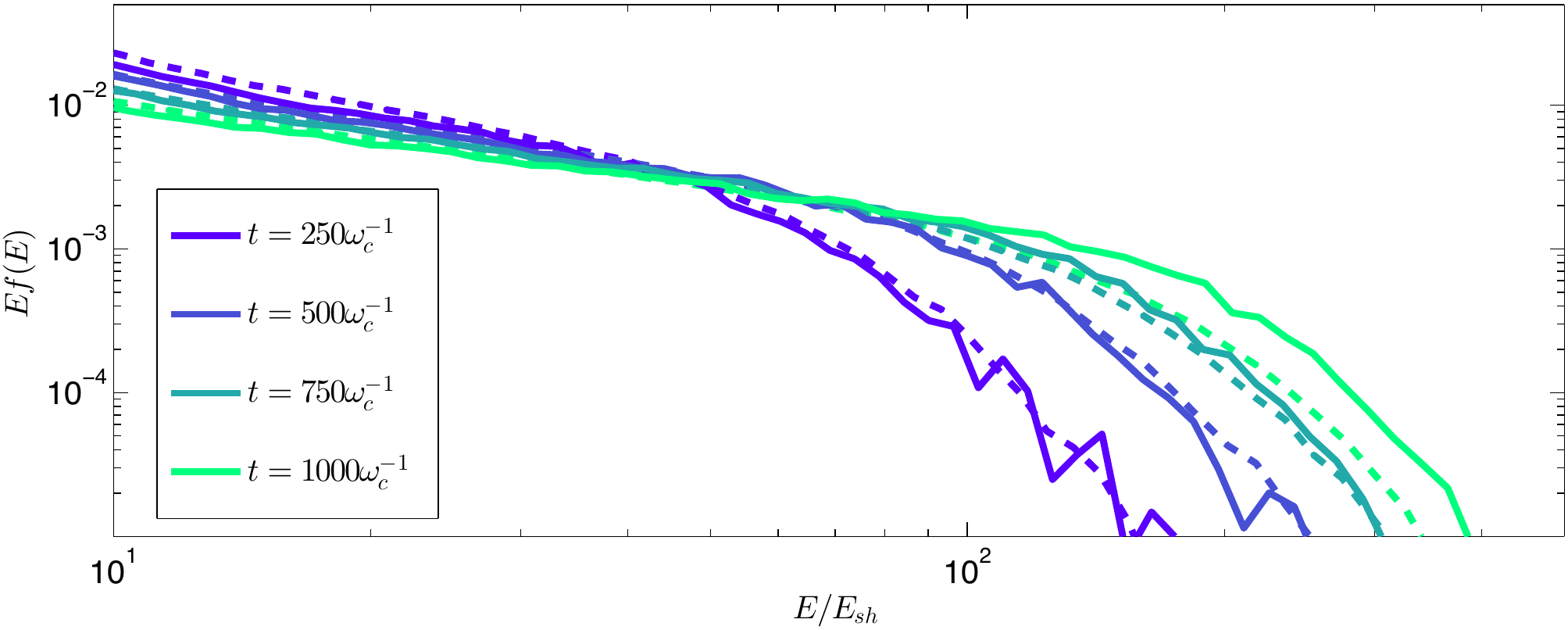}
\caption{\label{fig:AB}
Comparison of runs with different transverse size for $M=20$ parallel shock (Run A is 5 times larger than Run B, see Table \ref{tab:box}). 
\emph{Left panel}: Magnetic field profiles, as in the legend; 
$\langle B\rangle$ and $\max{B}$ correspond to the average over $y$ and the maximum of $B_{tot}(x,y)$, as a function of $x$.
$\max{B}$ in Run B is not shown since it almost coincides with $\langle B\rangle$. 
The averaged profiles are quite similar in the two runs, even if the spread from the mean value may locally be quite large in Run A. 
\emph{Right panel}: time evolution of non-thermal spectra for Run A (dashed lines) and B (solid lines).
Spectra agree very well until $t\approx 800\omega_c^{-1}$, after which the diffusion length at $E_{max}$ becomes comparable with the box size in Run A. 
\emph{A color figure is available in the online journal.}}
\end{figure}

A few comments on the role of the transverse box size are needed.
Figure \ref{fig:AB} shows the comparison of magnetic field profiles (left panel) and non-thermal ions' spectra (right panel) in Runs A and B, which have $L_y=1000c/\omega_p$ and $L_y=200c/\omega_p$, respectively (see Table \ref{tab:box}).
The left panel of figure \ref{fig:AB} shows the profile of the total magnetic field $B_{tot}=|{\bf B}|$, for both Run A and B.
The  upper curve illustrates the maximum value of $B_{tot}(x)$ for Run A, while bottom curves correspond to the average of $B_{tot}(x,y)$ over $y$, $\langle B_{tot}\rangle$. 
In Run B, $\max[B_{tot}](x)$ is almost indistinguishable from $\langle B_{tot}\rangle$, and is omitted in the plot.
It is important to notice that $\langle B_{tot}\rangle$ is almost independent of the actual transverse size of the box, as long as this is large enough to encompass the gyroradius of most of the ions.
In simulations with large transverse size, however, filamentation leads to prominent inhomogeneities, and the local field may vary significantly (compare the red and the magenta curves in figure \ref{fig:AB}).

The ion spectra obtained in Runs A and B are very similar to each other up to $t\lesssim 800\omega_c^{-1}$ (right panel of figure \ref{fig:AB}).
Beyond this time, the diffusion length of the highest-energy ions becomes comparable with the longitudinal box size in Run A, and $E_{max}$ does not increase any longer (left panel of figure \ref{fig:AB}). 
The DSA spectral slope is determined by the shock compression ratio only, but $E_{max}(t)$ is determined by the strength and the topology of magnetic irregularities that scatter the ions.
The fact that $E_{max}(t)$ does not depend on the transverse box size until $t\gtrsim 800\omega_c^{-1}$ implies that ---on average--- also the diffusion of accelerated ions is not strongly affected by filamentation.

We conclude that large transverse sizes are needed to capture the proper topology of the electromagnetic fields and the shock corrugation, which may have observational implications (see CS13).
However, moderately-large 2D simulations are still adequate to study important quantities in the long-term evolution of the shock, as the averaged level of magnetic field amplification, thereby returning realistic ion power-law distributions.   

\vspace{1mm}

\bibliographystyle{yahapj}
\bibliography{MFA}
\end{document}